\documentclass[fleqn,usenatbib]{mnras}

\usepackage{newtxtext}

\usepackage[T1]{fontenc}

\usepackage{graphicx}	% Including figure files
\usepackage{amsmath}	% Advanced maths commands
\usepackage{amssymb}	% Extra maths symbols
\usepackage[utf8]{inputenc}  
\usepackage[T1]{fontenc}
\usepackage{mathrsfs}
\usepackage{amssymb,gensymb}
\usepackage{multirow,enumerate,verbatim}
\usepackage{comment,hyperref}
\usepackage{color}
\usepackage{acronym}
\usepackage{xspace}
\usepackage[normalem]{ulem}
\usepackage{mathtools}
\usepackage{subfigure}% <-- changed

\usepackage[ruled,vlined,linesnumbered]{algorithm2e} 
\usepackage{makecell}

\SetKwInOut{Parameters}{Parameters}
\title{Neural likelihood estimators for flexible Gravitational wave data analysis}

\author[]{
Luca Negri,$^{1,2}$\thanks{E-mail:l.negri@uu.nl}
Anuradha Samajdar,$^{1,2}$
\\
$^{1}$Institute for Gravitational and Subatomic Physics (GRASP), Utrecht University, 3584 CC, Utrecht, The Netherlands \\
$^{2}$Nikhef, 1098 XG, Amsterdam, Netherlands
}

\date{Accepted XXX. Received YYY; in original form ZZZ}

\pubyear{2025}

\begin{document}
\maketitle

\begin{abstract}
In this paper, we develop a Neural Likelihood Estimator and apply it to analyse real gravitational-wave (GW) data for the first time. We assess the usability of neural likelihood for GW parameter estimation and report the parameter space where neural likelihood performs as a robust estimator to output posterior probability distributions using modest computational resources. In addition, we demonstrate that the trained Neural likelihood can also be used in further analysis, enabling us to obtain the evidence corresponding to a hypothesis, making our method a complete tool for parameter estimation. Particularly, our method requires around 100 times fewer likelihood evaluations than standard Bayesian algorithms to infer properties of a GW signal from a binary black hole system as observed by current generation ground-based detectors. The fairly simple neural network architecture chosen makes for cheap training, which allows our method to be used on-the-fly without the need for special hardware and ensures our method is flexible to use any waveform model, noise model, or prior. We show results from simulations as well as results from \texttt{GW150914} as proof of the effectiveness of our algorithm.
\end{abstract}

\section{Introduction}

The first observation of gravitational waves (GWs) in 2015 (\cite{LIGOScientific:2016aoc}) opened a new window into the universe. Since then, more than $\sim 200$ GW signals have been confidently detected by the LIGO \cite{LIGOScientific:2014pky}
 Virgo (\cite{VIRGO:2014yos}) Kagra (\cite{Kagra:2013}) (LVK) collaboration in the fourth observing run alone (\cite{gracedb}). As the number and complexity of GW events continue to grow (\cite{Abbott:2019,abbot:2021,Abbott_2023,abac:2025}), the computational burden of parameter estimation (PE) has become increasingly apparent (\cite{smith:2019}). In addition to the cost of analysing multiple events, tests of fundamental physics, such as probing deviations from General Relativity (\cite{LIGOScientific:2016lio,LIGOScientific:2019fpa,LIGOScientific:2021sio,LIGOScientific:2020tif}), require several additional parameters and repeated analyses per event. Waveform models incorporating richer physics (e.g., eccentricity or tidal effects) further increase the dimensionality of the parameter space, making traditional Bayesian inference very expensive. Re-analysis, whether to apply refinements or under different prior settings often requires starting from scratch, making previous analyses effectively obsolete. 

All signals observed in GWs so far have been Compact Binary Coalescences (CBCs) characterised by 15-17 parameters. PE proceeds by using Bayesian inference to compute the posterior probability distribution functions (PDFs) of the parameters of interest (\cite{Veitch:2008ur},\cite{Veitch:2009hd}) and typically uses a stochastic sampling algorithm to explore a high-dimensional parameter space. In addition to posterior PDFs, a component of Bayesian inference is the \emph{evidence}, a multidimensional integral over the product of the likelihood and the prior PDFs, signifying the support of a single hypothesis. This quantity is vital for hypothesis ranking (\cite{Veitch:2008wd}), including studies of the neutron star equation-of-state (\cite{LIGOScientific:2019eut}) and ranking different theories of gravity (\cite{LIGOScientific:2016lio}). The likelihood function is computed using GW data and, in a typical PE algorithm, is computed about $\mathcal{O}(10^6-10^8)$ times, making the likelihood computation the most costly part of the inference process. Due to its promising speed and robustness, machine learning (ML) has become a powerful tool in the last few years in every field of GW data analysis. A comprehensive review of ML-related works in GW astronomy can be found in Ref.~\cite{Cuoco:2024cdk}.\\

 In the recent past, particular strides have been made in the direction of faster and cheaper inference through simulation-based inference, and specifically, neural posterior estimation. Many of these methods perform GW analyses within the framework of a \emph{likelihood-free} inference both for CBCc (\cite{Green:2020hst,Dax:2021tsq,Dax:2022pxd,Dax:2024mcn,gupte2024,chua:2020,Gabbard:2021,Chatterjee:2024,kolmus:2024,uddipta:2024,Hu:2025}) and other sources (\cite{alvey:2024,desanti:2024}) enabling the generation of parameter PDFs on short timescales. \cite{Williams:2021qyt,Wong:2023,perret:2025} are other promising avenues to perform PE on GW signals using ML-based approaches where knowledge of the likelihood function is retained. Refs.~\cite{Green:2020hst,Dax:2021tsq,Dax:2024mcn} use neural posterior estimation by implementing \emph{normalizing flows} to perform rapid PE and output PDFs of parameters. The method requires extensive pre-training taking $\sim \mathcal{O}(10)$ days on a single NVIDIA A100 (\cite{Dax:2021tsq}). Once trained, inference can be performed in a matter of seconds. In \cite{Dax:2022pxd}, a method is proposed to further improve robustness by means of neural importance sampling, requiring $10^5$ draws from the neural posterior and compared to the true likelihood. Refs \cite{ashton:2021} and \cite{,Wong:2023,Wouters:2024oxj,Polanska:2024arc} integrate normalizing flows within a stochastic Markov Chain Monte Carlo (MCMC) sampler for an efficient jump proposal, reducing the number of likelihood evaluations necessary, the latter method, trained on-the-fly, also leverages gradient-based sampling and hardware acceleration by using Graphical Processing Unit (GPUs) and Tensor Processing Units (TPUs). Further, waveforms used are written in JAX (\cite{47008}), making the method extremely fast but relying on the availability of sophisticated computational resources as well as tailored waveform models. Ref \cite{perret:2025} uses Hamiltonian Monte Carlo to speed up PE for binary neutron stars by learning the gradient of the likelihood function with a deep neural network trained on-the-fly, achieving great reductions in computing times.  Refs.~\cite{Williams:2021qyt,Williams:2023ppp,prathaban:2024} incorporate normalizing flows and $\beta$-flows within the nested sampling (\cite{skilling:2006}) algorithm, using them to guide live points toward higher-likelihood regions, which also allows to reduce the number of likelihood evaluations. This approach enables direct computation of the evidence, avoiding the need for post-processing. Another approach to speedup the analysis is to directly reduce the computational costs of evaluating the likelihood function, and alongside many analytical methods, (\cite{barak:2018,narola:2023,Vinciguerra:2017,Morras:2023,canizares:2015}), ML-based solutions have also been proposed. Ref.~\cite{Graff:2011gv} introduced the concept of Neural Likelihood Estimators (NLE) trained on-the-fly to approximate the true likelihood of GW-related problems, and showed great promise in low-dimensional scenarios. Neural likelihood estimators have been explored more recently in \cite{papamakarios:2019} and for LISA data analysis in \cite{Martin_Vilchez:2025} and \cite{gammal:2025}, which uses Gaussian process interpolation as an approximant for the likelihood. Evaluating the likelihood on a grid in intrinsic parameter space and using Gaussian processes to directly compute the marginalised likelihood has also been in use for inference with particularly expensive waveform models for real gravitational wave signals (\cite{Lange:2018pyp,wagner:2025,Williams:2020}).  Our algorithm follows and expands on these ideas, keeping the conceptually simple nature of the approach in ~\cite{Graff:2011gv}, implementing recent advances in machine learning techniques, and extending the method to be able to perform inference on real GW signals.

In this work, we present a machine learning-based PE method that retains access to the true likelihood function by generating an estimator on the fly, enabling direct estimation of the posterior distributions and Bayesian evidence, considerably reducing computational costs when compared to standard sampling methods. We require around $10^5$ true likelihood evaluations, about 10-100 times less than standard PE. Our approach uses a compact, fully-connected residual network trained during the sampling process itself. This makes our method straightforward to implement and flexible to adapt to any likelihood function. The total computational costs, including training, on a single CPU are on the order of tens of minutes for a single binary black hole (BBH) event. We refer to our algorithm as \texttt{FLEX} and in this paper, we outline a proof-of-principle study highlighting its advantages and limitations. In this work, we focus on relatively high-mass BBH systems. 

In Sec.~\ref{sec:method}, we discuss the methods used to develop our algorithm, including a summary of Bayesian inference and details of our neural network architecture and the way it is trained. Sec.~\ref{sec:results} shows the results of validating our neural likelihood algorithm and applications to simulated data as well as the real signal \texttt{GW150914} (\cite{LIGOScientific:2016aoc}). We summarise and discuss the limitations of our methodology in Sec.~\ref{sec:discussion} and conclude in Sec.~\ref{sec:conlcusions}.

\section{Method}
\label{sec:method}
In this Section, we describe the methodology used to implement our algorithm. We start with a general description of Bayesian inference and go on to describe individual components of our new approach to incorporate a neural likelihood estimator (NLE) in a standard Bayesian inference algorithm. 

\begin{algorithm}
    \caption{\texttt{FLEX} pseudocode}
    \label{tab:algo}
    % \SetAlgoLined
    \KwData{Prior distribution $\pi(\theta)$, true likelihood function $\mathcal{L}(\theta)$}
    \Parameters{Number of samples per temperature $Nt$, temperature ladder $T_{ladder}$,maximum number of cycles, $ESS$ threshold }
    \KwResult{Posterior samples $\mathcal{P}$, neural likelihood $NN(\theta)$}
    $ \Theta = [\vec\theta_0, ..\vec\theta_k..,\vec\theta_{Nt}] \text{ with }  \vec\theta_k \sim \pi $ \\

\textit{    Annealed-KDE algorithm to obtain training samples}:

    \For{T in $T_{\mathrm{ladder}}$}{
        $\mathbf{v} \leftarrow \mathcal{L}(\Theta)^{1/T}$ \\
        $\Theta_T \leftarrow [\theta_0, ..\vec\theta_k.. ,\theta_{Nt}] \text{ with }  \vec\theta_k \stackrel{i.d.d.}{\sim} KDE(\Theta,\mathbf{v})$\\
        $\Theta \leftarrow [\Theta,\Theta_T]$
    }
    
    \textit{Obtaining posterior and approximant with FLEX:}
    
    \For{\text{cycle} in \text{max cycles}}{
        $NN \leftarrow trainNN(\Theta,\mathcal{L}(\Theta))$\\
        $\mathcal{P} \leftarrow \text{MCMC}(\pi,NN)$\\
        $ESS \leftarrow computeESS(NN(\mathcal{P}),\mathcal{L}(\mathcal{P}) )$\\
        \eIf{$ESS<$threshold}{
            $\Theta \leftarrow [\Theta,\mathcal{P}]$
        }{
        return $\mathcal{P}, NN$
        }   
    }
\end{algorithm}

\subsection{Bayesian inference}
In a Bayesian framework, all information about the parameters 
of interest is encoded in the posterior probability density function (PDF), given by Bayes' theorem:
\begin{equation}
 p(\vec{\theta}|\mathcal{H}_s,d) 
 = \frac{\mathcal{L}(d|\vec{\theta},\mathcal{H}_s)\,p(\vec{\theta}|\mathcal{H}_s)}{\mathcal{Z}},
 \label{eqn:Bayes}
\end{equation}
where $\vec{\theta}$ is the set of parameter values and $\mathcal{H}_s$ is the hypothesis that a 
GW signal depending on the parameters 
$\vec{\theta}$ is present in the data $d$ (\cite{Veitch:2009hd},\cite{veitch:2015}). 
For parameter estimation purposes, the factor $\mathcal{Z}$, called the \emph{evidence} 
for the hypothesis  $\mathcal{H}_s$, is effectively set by the requirement that PDFs are normalised. 
Assuming the noise to be 
Gaussian, the \emph{likelihood} $\mathcal{L}(d|\vec{\theta},\mathcal{H}_s)$ of obtaining data $d(t)$ given the 
presence of a signal $h(t)$ is determined by the proportionality
\begin{equation}
 \mathcal{L}(d|\vec{\theta},\mathcal{H}_s) \propto \exp{\left [-\frac{1}{2}\langle d-h(\vec{\theta})|d-h(\vec{\theta}) \rangle\right ]},
 \label{eqn:lhood}
\end{equation}
where the noise-weighted inner product $\langle\cdot\,|\,\cdot \rangle$ is defined as (\cite{Cutler:1994ys})
\begin{equation}
\langle a|b \rangle = 4\Re \int_{f_{\rm low}}^{f_{\rm high}} \frac{\tilde{a}^\ast(f)\,\tilde{b}(f)}{S_h(f)}\,df.
\end{equation}
Here, a tilde refers to the Fourier transform, and $S_h(f)$ is the noise power spectral density (PSD). The evidence of the signal hypothesis $\mathcal{H}_s$ is given by the following integral over the full parameter space $\vec{\theta}$:
\begin{equation}
    \mathcal{Z} = \int_{\vec{\theta} }\mathcal{L} (d|\vec{\theta},\mathcal{H}_s) p(\vec{\theta}|\mathcal{H}_s) d\theta.
\end{equation}

For a GW signal in frequency domain denoted by $h(f)$, the optimal signal-to-noise-ratio (SNR) is given by
\begin{equation}
    \rho^2 = 4 \Re \int_{f_{\rm low}}^{f_{\rm high}} \frac{|h(f)|^2}{S_h(f)} df.
\end{equation}
Over a network of detectors, the network SNR is then given by the quadrature summation $\sqrt{\Sigma_{i=1}^I \rho_i^2} $ for $I$ detectors.\\

In GW parameters estimation, evaluating Eqn. \ref{eqn:lhood} is the slowest and most computationally expensive task. For this reason, we aim to approximate it with a neural network which can be sevaral orders of magnitude faster per evaluation. Unless specified, whenever we mention the number of true likelihood evaluations, we are referring to the number of times Eqn. \ref{eqn:lhood} has been solved analytically, rather than through our neural network. Since it is so cheap to evaluate, the total number of neural network evaluations will have little impact on the total computational costs. 

 \subsubsection{Gravitational waves likelihood parameterisation}
    \label{subsec:params}
    In Eqn.~\ref{eqn:Bayes}, $\vec{\theta}$ refers to the parameter set describing a CBC signal; a BBH signal is typically characterised by 15 parameters. In our case however, we have a total of 9 parameters. Throughout our analyses, we sample on the following parameter set:
    \begin{equation}
       \vec{\theta} = \{\mathcal{M},q,\chi_1,\chi_2,\theta_{jn},\psi,\kappa, \epsilon,t_{\mathrm{det}}\}.
    \end{equation}
    For a CBC system with component masses $m_1$ and $m_2$, $\mathcal{M}$ is the chirpmass defined as:
    \begin{equation}
        \mathcal{M} = M \eta^{3/5},
    \end{equation}
    where $M=m_1+m_2$ is the total mass of the binary and $\eta = \frac{m_1 m_2}{M^2}$ is the symmetric mass-ratio. The sampling parameter $q=m_2/m_1$ is the mass-ratio of the binary.
    The dimensionless spin parameter of each companion mass $m_i$ is defined as
    \begin{equation}
        \vec{\chi_i} = \frac{\vec{S}_i }{m_i^2},
    \end{equation}
    where $\vec{S}_i$ is the spin vector of the $i^{\rm th}$ object and $\chi_i = \vec{\chi_i} \cdot \hat{L}$ is the spin component along the orbital angular momentum of the binary. For simplicity, in our work we have chosen spins aligned with the orbital angular momentum, in other words, assuming the observer is located along the $\hat{z}$ axis of a binary's orbital plane, only the $z$ components of the spins survive. 
    We follow the alternate sky location and time parameterisation introduced in~\cite{Romero-Shaw:2020owr} to improve sampling efficiency. Instead of sampling in equatorial coordinates right-ascension and declination ($\alpha, \delta$) and geocentric coalescence time $t_c$, we sample in terms of the signal's arrival time at a single detector and sky location relative to the detector baseline, using the zenith and azimuthal angles $(\kappa, \epsilon)$. This re-parameterisation aligns the sampling axes with the ring-shaped likelihood structure induced by time-of-flight delays, reducing parameter correlations and accelerating convergence. The declination, right ascension, and time at geocenter parameters used typically in astrophysics can be obtained by a change of coordinates after sampling the posterior. A GW waveform of length $T$ at a reference time $t_c$ and frequency-bin $j$ can be written as
    \begin{equation}
        h_j = h_j(t_c) \exp{\left[-2\pi i j \frac{(t - t_c)}{T}\right]}.
    \end{equation}
    In case of the waveform consisting only of the dominant mode, if the waveform is known at a reference phase $\phi_c$ and a reference luminosity distance $D_0$, at an arbitrary phase and distance, the waveform may be written respectively as (\cite{thrane:2019})
    \begin{equation}
        h(\phi_c) = h(\phi_c=0) \exp{(2i\phi_c)}
    \end{equation}
    and
    \begin{equation}
        h_j(D_L) = h_j(D_0) \left( \frac{D_0}{D_L} \right).
    \end{equation}
    Using the parametrisation above, the extrinsic parameters $\phi_c$ and $D_L$ may be marginalised analytically and numerically from a lookup table, respectively. Finally, our parameter set is reduced to 9 dimensions in total.  

 \begin{figure}
            \centering
            \includegraphics[width=0.45\textwidth]{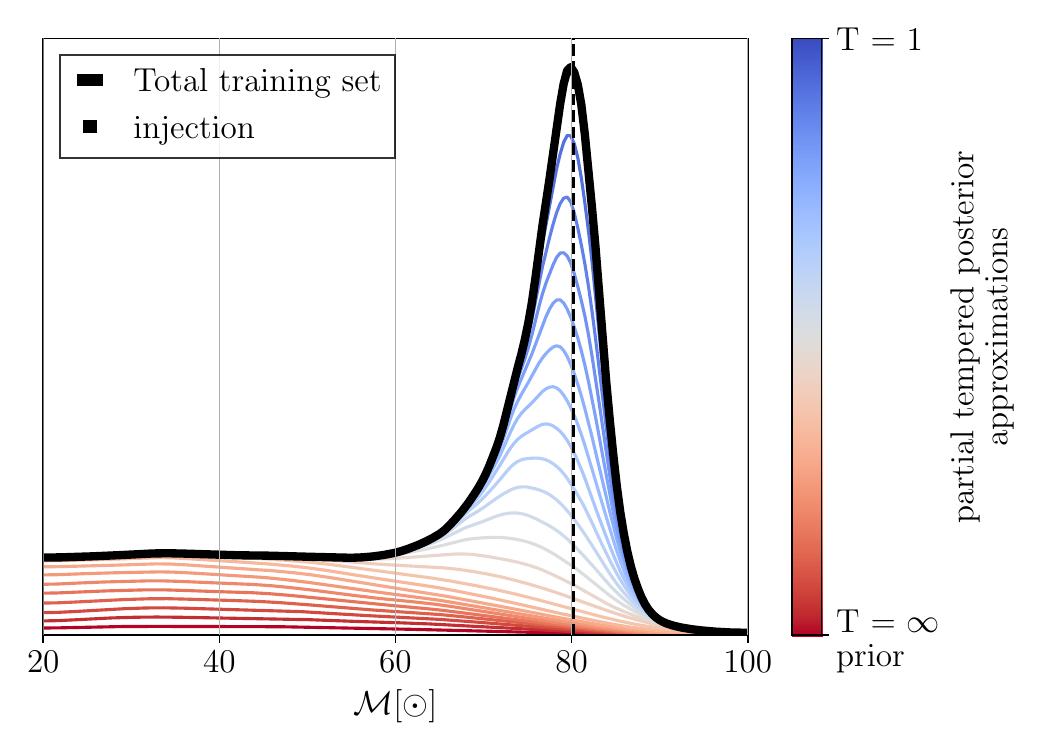}
            \caption{Plot showing how the training set is constructed. Starting from the prior, the posterior is iteratively approximated by a tempered KDE. Sampling from a tempering schedule from $T=\infty$ (prior) to $T = 1$ (posterior), we add new training samples from each successive KDE. The colored lines indicate the total number of samples obtained before that temperature.}
            \label{fig:prior_vs_kde}
        \end{figure}
    \subsection{Generating training samples}
    \label{subsec:training_samples}
        The first step of the \texttt{FLEX} algorithm is also the most delicate: training samples have the fundamental task of representing the real likelihood as best and efficiently as possible; maximizing precision with low computational costs boils down to employing a smart sampling scheme. The full likelihood function $\mathcal{L}(\vec{\theta}|\mathcal{H},d)$ depends on the data $d$, the signal plus noise hypothesis $\mathcal{H}_s$ and the signal parameters $\vec{\theta}$. We aim to train the NLE on the fly for every new analysis, so the data and the hypothesis remain fixed. The likelihood value now depends only on $\vec{\theta}$, so the training set will be a set of points $[\vec\theta_i]$ and the corresponding likelihood $\mathcal{L}_i = \mathcal{L}(\vec\theta_i|d,\mathcal{H}_s)$. The NLE acts as an interpolator between these points.\\

        The cost of generating the training set depends both on the cost of evaluating the likelihood functions and the total number of points required. For this algorithm to be faster, this number must be kept below the values used for standard PE, which range around $10^6-10^8$. We aim to keep the number of samples needed to train \texttt{FLEX} around $\mathcal{O}(10^5)$, to ensure a speedup factor of at least 10 times.\\
        
        It is not easy to represent the $\mathcal{O}(10)$ dimensional parameter space with such a small number of samples, and since the stochastic sampler will probe the whole prior range, the NLE needs to be accurate over all of it. At the same time, the posterior points will only come from the small-volume high-likelihood regions (around a billionth of the prior volume in our analysis). The resolution of the training set in this region needs to be quite high to ensure a reliable estimate of the posterior. This means that our training samples need to roughly approximate the final posterior already. We developed a novel algorithm that can obtain a good approximation of the posterior with a fixed number of likelihood evaluations, as well as samples from the neighboring regions.\\
        
        Introducing a temperature parameter $T$, it is possible to define a tempered version of the posterior
        \begin{equation}
            p_T(\vec{\theta}|d)
        \propto p(d|\vec{\theta})^{1/T} p(\vec{\theta}).
         \label{eqn:tempered_post}
        \end{equation}
        For $T=1$, Eqn. \ref{eqn:tempered_post} returns the standard posterior, while for $T = \infty$ we obtain the prior. It is now possible to smoothly interpolate between the prior and the posterior by choosing different values of $T$.\\
        Sequentially sampling posteriors with different $T$s would allow a balanced generation of samples in volumes of equal interest for neural likelihood: sparse samples in areas of low likelihood, characterising the bulk of parameter space, and an increasing density around the posterior to help improve the accuracy. Of course, since we do not have access to the real posterior, we must use an approximation. We have developed a method based on this idea to generate training samples, which we call annealed Kernel Density Estimate (KDE) (\cite{kde_1,kde_2}). Starting from the prior, the posterior is iteratively approximated by a tempered version of the KDE, by giving each sample a weight $w\propto \mathcal{L}(\vec{\theta})^{1/T}$. A new set of samples is drawn from the KDE, and, after their likelihood values are obtained, they are added to the total sample pool. A new KDE is then calculated by computing the weights with a lower temperature, and the cycle continues. The final samples obtained will be our first posterior approximation. An example of how the construction of the training set is performed is presented in Fig. \ref{fig:prior_vs_kde}.\\

       Methods to generate the training set would need samples from the highest-likelihood region as well as the bulk of parameter space. Examples of other methods might be to use the intermediate samples obtained from an optimization algorithm, like Differential evolution (DE) (\cite{Storn:1997uea}). We found that using samples from DE performed similarly to those from the annealed-KDE. We believe that our method is quite robust to different choices in the distribution of initial samples, as long as the two conditions listed above are met. In future works, we plan to expand on this and additional methods of generating initial samples. In this paper, all initial training samples are obtained through the annealed-KDE method.
        
    \subsection{Neural Likelihood}
    \label{subsec:network}

         One of the properties that made neural networks ever-present in the machine learning literature is their ability to approximate any non-pathological mapping $\mathbb{R}^n \xrightarrow{}\mathbb{R}^m$, if allowed to have at least one hidden layer with an arbitrarily large number of neurons with a non-linear activation function. This makes them universal approximators. By increasing the number of neurons, their expressiveness can increase quite quickly.\\
         As mentioned in section \ref{subsec:training_samples} in this problem setup, the NLE will have to learn a mapping from the parameter space to the likelihood space.   If we set $n$ to be the dimensionality of $\vec{\theta}$ and $m$ to be equal to 1, it is then possible to find a parametrization $\Phi$ for a Neural Network ($NN$) such that 
         \begin{equation}
             NN_{\Phi}(\vec{\theta}) \sim \mathcal{L}(\vec{\theta}) 
         \end{equation}
         
         $NN_{\Phi}$ can then be used in any stochastic sampler to find the posterior, and if the evaluation of $NN_\Phi(\vec\theta)$ is faster than $\mathcal{L}(\vec\theta)$, the posterior will also be obtained much faster. The rest of this section will be dedicated to explaining how the \texttt{FLEX} framework utilises the samples obtained in Sec. \ref{subsec:training_samples} to train a Neural Network to approximate the likelihood function.\\
        
        Each training sample $\vec{\theta}_i$ is pre-processed before being passed to the network by means of normalization. The angular variables are passed to the network as their sine and cosine; this will make the output of the network periodic in this dimension. Parameters that have a clear periodicity of half of the full period will have the sine and cosine of $2\theta$ passed as well. Other parameters are scaled to have a mean of 0 and a variance of 1. The likelihoods are always passed as a natural logarithm $log\mathcal{L}_i$, and the median (instead of the mean) in log-space is chosen for normalization. The output of the network is then unnormalised to perform inference. Whenever $NN(\vec\theta)$ is mentioned, the normalization and unnormalisation operations are implied. \\

        To ensure that $NN_\Phi(\vec\theta)$ remains fast to evaluate, the size of the network must be kept relatively small. This will speed up both training and inference. A small network also reduces the chances of overfitting. In our use case, overfitting will take the form of spurious modes appearing in the posterior. Borrowing terminology from Large Language models, it is as if the NLE hallucinates a posterior mode in regions that are not supported by the training set.  The final network we chose is a 4-layer deep and 64-node wide ResNet architecture. Together with the input layer and the single node output layer, this yields a total of $\sim 15$k trainable parameters. The Gaussian Error Linear Units (GELU) activation function was chosen for the hidden layers.\\
        
        The loss function is composed by two terms:

        \begin{align}
            Loss &= L_{MSE} + \lambda L_{R}
        \end{align}
        Which take the form:
        \begin{align}
            L_{MSE} &= \frac{1}{N}\sum_{i=0}^{N} ( e^{\log\mathcal{L} ({\vec\theta_i)}} -  e^{\mathrm{NN}_\Phi({\vec\theta_i)}} )^2v_i\\
            L_{R} &= \|\vec{\Phi}\|_1
        \end{align}

        Where $N$ is the size of the training set. $L_{MSE}$, measures the accuracy of the network with a weighted Mean Squared Error(MSE) between the real and predicted exponential of the log-likelihood. The exponential ensures that contributions to $L_{MSE}$ will mostly come from the samples associated with the highest likelihoods. $NN(\vec{\theta})$ will be more accurate in the regions where we expect the bulk of the posterior to lie, while larger errors are allowed in the remaining parameter space. However, these low-likelihood regions are also undersampled, and the risk of spurious peaks appearing as a consequence of overfitting needs to be taken into account. To balance the difference in resolution in parameter space, a weight term $v_i$ is associated to every sample. The idea is to give more weight to the samples that lie in sparsely populated areas of parameter space. A KDE is computed over the whole training set, and the corresponding density is computed for each sample. Finally, $v_i = \sqrt[d]{KDE({\vec\theta_i})}$ where $d$ is the parameter space dimensionality. The weight will be proportional to the mean distance between nearest-neighbours around the sample. The regularization term $L_{R}$ penalises overfitting even further by computing the $L_1$ norm over the network parameters $\Phi$. The $\lambda_{R}$ hyperparameter balances between the two losses. For the analysis in this paper we set  $\lambda_{R} = 10^{-6}$ .\\

         Since the network must be retrained for every new signal, we must adopt a stable and flexible training scheme. Adamax (\cite{kingma:2017}) was chosen as the optimiser, and, as the learning rate scheduler, the cosine annealing with warm restarts (\cite{loshchilov:2017}) was deemed the best option, due to its self-stabilising nature. A gradient clipping algorithm (\cite{zhang:2020}) with an adaptive threshold is also used to further stabilise the training procedure, by reducing the incidence of sudden jumps around the $NN$ parameter space caused by the exploding gradient of the loss function.

    \subsection{Markov Chain Monte Carlo}
        \label{subsec:mcmc}
        Once the neural likelihood is trained, we can use it in any stochastic sampler to generate the posterior distribution. Any sampler can be run on the \texttt{FLEX} NLE, but to get the best out of the framework, it must be capable of both handling the complexities of the gravitational waves likelihood, such as multimodalities and non-Gaussianities, and exploiting the computational advantages brought by the neural network.\\
        In our current framework, the call to the neural network function has relatively large overhead per call of $\mathcal{O}$(ms), which is comparable to the time taken by the true BBH likelihood: if the neural likelihood is naively implemented in a stochastic sampler, it would negate any speed advantage. If the code is vectorised, a single neural likelihood call can process a large number of samples at once, and the overhead time becomes negligible. To gain the speedups of the neural likelihood, we thus need a sampling framework that is highly parallelisable so that vectorisation is possible.  We decided to use a Markov Chain Monte Carlo (MCMC) sampler with a large number of walkers; each walker explores the likelihood surface independently, making the call to the likelihood function trivially parallelisable.\\
        
        Common MCMC techniques (\cite{Hogg:2017akh,Sharma:2017wfu}) for gravitational waves implement affine-invariant transformations and parallel tempering (\cite{Gilks01091998}). Affine invariant transformation sampler use what is commonly referred to as a "stretch" move as the proposal for new points. The stretch move selects a random point from an ensemble, and proposes a jump of random length in that direction. This transforms a complex multidimensional distribution into an easier one-dimensional one. The sampler can then handle non-Gaussian distributions, which are typical for GW posteriors. Parallel tempering uses many chains of walkers in parallel at different posterior temperatures. The higher temperature posteriors are flatter, as shown in Sec \ref{subsec:training_samples}, and this helps walkers jump from one posterior peak to another, avoiding mode collapse.\\
        
        We decided to use the \texttt{eryn} MCMC sampler (\cite{Karnesis:2023ras}) to recover the posterior from the NLE. It implements likelihood vectorisation, parallel tempering, and affine transformations.
        Internal testing showed that for the range of signals analysed in this paper, the \texttt{eryn} posterior closely matched those obtained through the \texttt{dynesty} sampler (\cite{speagle:2020}), which is the one that is more widely used in this field. We will further address different sampler choices in Sec \ref{subsec:followup}   
    
    \subsection{Assess results and retrain}
        \label{subsec:assess}
    
        Finally, we need to assess the accuracy of the posterior distributions we obtain. The posterior found by \texttt{FLEX} might lie in a region of parameter space where there is not much support from the training set, and this could potentially lead to large errors in the neural likelihood. A standard method to assess the accuracy of a posterior obtained through an approximate likelihood is to compute the number of effective posterior samples (\cite{Kong1992}).  Defining a weight $w_i$ for each posterior sample as the ratio between the true likelihood and the approximate likelihood
        \begin{equation}
             w_i = \mathcal{L}(\theta_i)/\mathrm{NN} (\theta_i),
        \end{equation}
        The number of effective samples, or the effective sample size (ESS) will be: 
        \begin{align} 
  {\rm ESS} = \frac{(\sum_{i} w_i)^2}{\sum_{i} w_i^2}.
        \end{align}

        The weights are normalised so that the largest ratio over the whole set of posterior samples is equal to 1.  Only if the two likelihoods agree, up to a constant multiplicative factor, over all of the posterior points, the weights will be all close to 1.\\
    
        If the $ESS$ is below a certain threshold, we reject the posterior, and a new training cycle will be triggered to improve the accuracy of the neural likelihood in that region of parameter space. This is achieved by adding the posterior points and their true likelihood evaluation to the training set. Moreover, samples from the tempered MCMC chains are added. Since the higher temperature posteriors are considerably wider, they give support to areas around the primary mode of the rejected posterior, but also have a higher chance of sampling secondary modes. This also helps to avoid the problem of mode collapse, a situation in which the neural likelihood focuses only on one mode of the posterior, ignoring potentially more interesting secondary modes. In this work, the temperature chosen for the retraining samples was selected by trial and error. Ref. \cite{Saleh:2024} proposes a method to obtain the optimal temperature to maximise the $ESS$. In follow-up work, we plan to implement this method to improve the quality of the retraining samples.  \\

        The algorithm will restart training cycles until a posterior is accepted. The maximum number of training cycles can be set by the user. For this work the maximum number of cycles was kept to 6, and the number of samples added to the training set per cycle is $2\times10^4$. Since the first phase already adds $10^5$ samples, the maximum number of true likelihood evaluations for this setup is $2\times10^5$.

    \subsection{Follow-up analysis: changing samplers}
        \label{subsec:followup}
        A unique advantage of this algorithm is having access to the fully trained neural likelihood after the posterior is obtained. This very fast approximation of the true likelihood can now be used to speed up any subsequent analysis. To put this to the test, we decided to add a final step to the algorithm and run inference on the trained NLE with a different sampler.\\
        
        Bayesian evidence is not directly accessible through MCMC samplers, while other algorithms, such as nested sampling (\cite{skilling:2006}) or Sequential Monte Carlo (SMC) (\cite{moral:2006}), can compute it directly. Standard nested sampling algorithms are notoriously hard to parallelise over a large number of threads, even though recent efforts have shown this capability ( \cite{smith:2019},\cite{yallup:2025}). Mass parallelisation with SMC is quite straightforward, allowing the use of efficient vectorisation, so we chose the latter.
        Starting from an arbitrary distribution (e.g., the prior), SMC algorithms iterate through a series of tempered posteriors in an annealing process, until a final posterior at temperature 1 is reached. A set of points is obtained from a high-temperature posterior, which is first resampled based on importance weights with respect to the next lower temperature in the ladder and subsequently perturbed through an MCMC process. The importance weights associated with the samples can be used to compute the evidence, and since walkers are evolved independently from each other this makes it easy to parallelise. For this task, we use the \texttt{pocomc} (\cite{Karamanis:2022}) sampler, additionally, it has been recently validated for use with real gravitational wave analysis (\cite{williams:2025}) and gives comparable results obtained by the widely used in GW data analysis \texttt{Dynesty} nested sampler.

\section{Results}
    \label{sec:results}
        \begin{figure}
        \centering
            \includegraphics[width=0.5\textwidth]{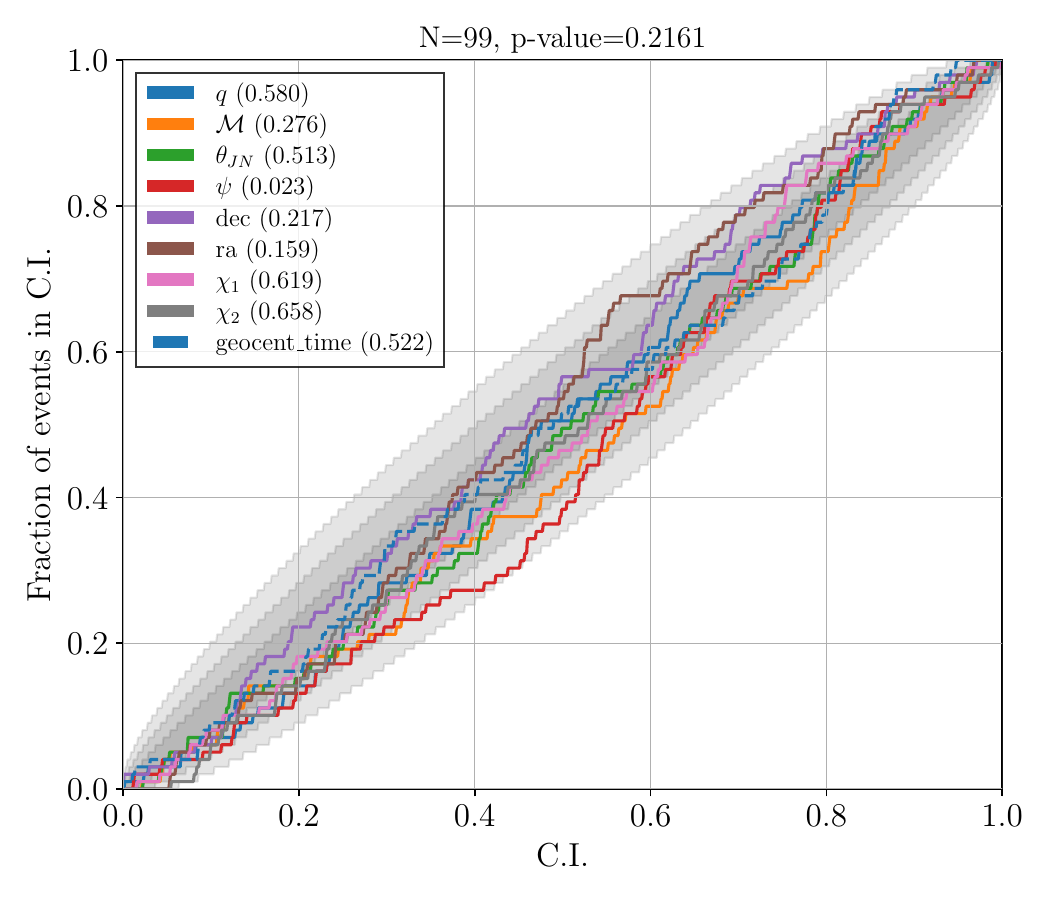}
            \caption{Plot showing the fraction of times each simulated value of the parameter falls within the same credible region. A perfect example would be a diagonal line with deviations worsening as the plots like within the shaded regions starting outwards from the diagonal line ($1\sigma, 2\sigma, 3\sigma$ respectively). Some deviations from the straight diagonal line are due to the finite number of sources. The final p-value howeve,r shows good agreement with our expectation.}
            \label{fig:pp}
        \end{figure}
    A robust NLE-based algorithm must satisfy the following criteria: (i) On average, lower computational requirements than a conventional Bayesian analysis, (ii) Final posterior PDFs statistically comparable to posterior PDFs from a standard Bayesian algorithm, and (iii) Pass diagnostic tests over a wide range of signal parameters. In the following, we look into each criterion in detail. We detail robustness tests by showing a probability-probability (PP) plot over many simulations, breaking down the computational costs of the algorithm and its convergence. Finally, we will apply our algorithm to analyse the real signal, \texttt{GW150914}, and compare results with standard analysis methods and further validate its robustness by performing the analysis with multiple waveform models. The true likelihood evaluations as well as post-processing of results have been carried out using the \texttt{bilby} package \cite{Ashton:2019} while the neural network architecture has been implemented in \texttt{pytorch} (\cite{PT}).
        
    \subsection{Injection studies}
        To assess the robustness of the \texttt{FLEX} algorithm, we first analysed simulated signals injected in Gaussian noise. While we look at the full distribution of results to validate the overall robustness, we focus on a single simulation to probe computational costs. 
    
        \subsubsection{PP-plot and testing robustness}
        \label{subsubsec:pp}
        
        \begin{table}
            \centering
            \begin{tabular}{|l|l|}
                \hline
                \textbf{Parameter} & \textbf{PDF} \\
                \hline
                 $q$& $\mathcal{U}[0.125,1]$ \\
                 $\mathcal{M} [\mathrm{M}_\odot]$& $\mathcal{U}$ in $m_1$ and $m_2$ $[20,100]$ \\
                 $\chi_1$& $\mathcal{U}[-1,1]$ \\
                 $\chi_2$& $\mathcal{U}[-1,1]$ \\
                 $D_L[\mathrm{Mpc}]$ & $\mathcal{U}^3 [10,5000]$\\
                 $\theta_{jn}$& $\cos[0,\pi]$ \\
                 $\psi$& $\mathcal{U}[0,\pi]$ \\
                 $\phi$& $\mathcal{U}[0,2\pi]$ \\
                 DEC& $\sin[-\pi/2,\pi/2]$ \\
                 RA &  $\mathcal{U}[0,2\pi]$\\
                 \hline
            \end{tabular}
            \caption{Prior ranges used both for the Bayesian analysis and to sample the parameters for the injections.}
            \label{tab:prior}
        \end{table}

        \begin{table}
            \centering
            \begin{tabular}{|l|c|}
                \hline
                \textbf{Parameter} & \textbf{value} \\
                \hline
                No. true likelihoods: &\\
                \hspace{0.2cm} (1st cycle) & $10^5$\\
                \hspace{0.2cm} (Tuning cycles) & $2\times 10^4$ \\
                \hline
                Max No. cycles & 6 \\
                \hline
                $N_{post}$ & 5000\\
                \hline
                $ESS/N_{post}$ threshold & 50\% \\
                \hline

            \end{tabular}
            \caption{Table reporting hyperparameters of the \texttt{FLEX} algorithm}
            \label{tab:hyperparameters}
        \end{table}
        
        To assess the statistical robustness of the results obtained by \texttt{FLEX}, we simulated 99 BBH systems in a 3-detector network of LIGO Hanford-Livingston and Virgo (HLV) in Gaussian noise coloured with current O4 sensitivities (\cite{Capote:2024rmo,LIGO:2024kkz}). All signals were chosen to have SNRs $\in [12,30]$. Both injection and recovery have been performed with the IMRPHenomD (\cite{Husa:2016}) waveform model. The signal parameters were sampled randomly from the prior distributions show in Tab. \ref{tab:prior}. The hyperparameters of the \texttt{FLEX} algorithm were set to the ones in Tab. \ref{tab:hyperparameters},  are kept the same for this and every following analysis, unless specified. Fig.~\ref{fig:pp} is a PP plot, showing the fraction of times an injected value of a parameter for the ensemble of these injections falls within that specific credible interval. An ideal PP plot will return a diagonal line (50\% of the time the injected value should fall within the recovered 50\% credible interval); however, the finite number of sources can lead to some deviations away from the diagonal. To quantify such deviations, the plot shows, in order of moving away from the diagonal, widths of $1\sigma$, $2\sigma$, and $3\sigma$, respectively. All parameters fall within the $3\sigma$ bounds. The total P-value of 0.21 also shows that the results are statistically sound. For individual parameters, only the polarisation parameter $\psi$ has a p-value below the 0.05 threshold. Explicitly checking the posterior PDFs of $\psi$ from the individual simulations did not flag a systematic issue. 
        Further analysis will be carried out to check if this is just a statistical anomaly and to exclude the hypothesis of systematic errors introduced by the algorithm.\\   
    
\begin{figure}
            \centering
            \includegraphics[width=0.5\textwidth]{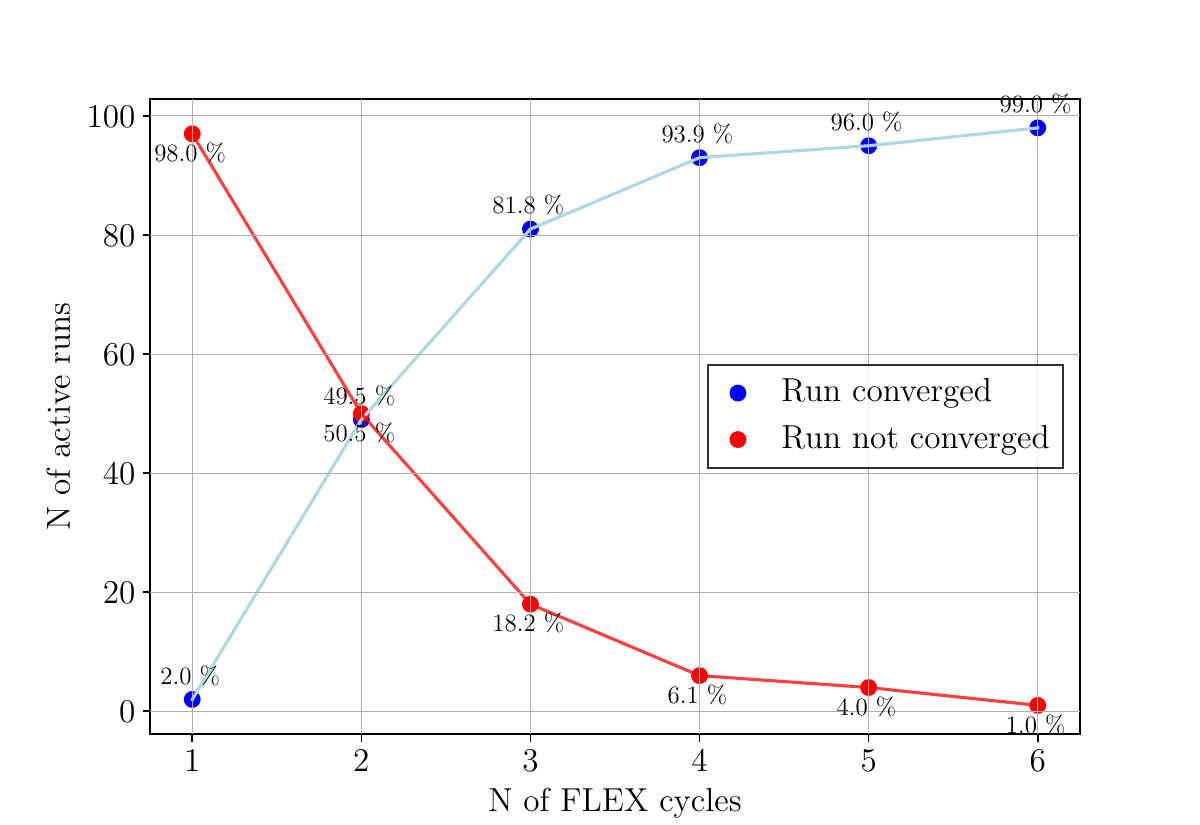}
            \caption{Percentage of runs which have converged as a function of \texttt{FLEX} cycles. A run is considered converged if the $ESS$ ratio is above 50\%. Out of the 99 injections, only 1 did not converge before the limit of 6 cycles. The signal parameters and the runs are the same as those used to obtain the p-p plot.}
            \label{fig:training_vs_SNR}
        \end{figure}

        For this same simulation set, we show the fraction of runs that have (not) converged before a certain training cycle in Fig.~\ref{fig:training_vs_SNR}. Two out of the 99 injections reached the effective sample size threshold after the first cycle, the median $ESS/N_{tot} = 5.7^{24.1}_{0.2}\%$ being well short of the 50\% threshold (\emph{c.f.} Sec.~\ref{subsec:assess}). Going onto the second cycle and adding samples from the first \texttt{FLEX} posterior dramatically increases the accuracy: now half of the runs have reached convergence at the end of the second cycle. Going further with the cycles, more and more runs reach convergence, and only 1 run failed to meet the criteria before the final cycle. This means that 99\% of the time \texttt{FLEX} reaches the 50\% $ESS$ ratio with less than or equal to $2\times 10^5$ likelihood evaluations.\\
        
\begin{figure}
        \centering
        \subfigure[]{
            \includegraphics[width=0.45\textwidth]{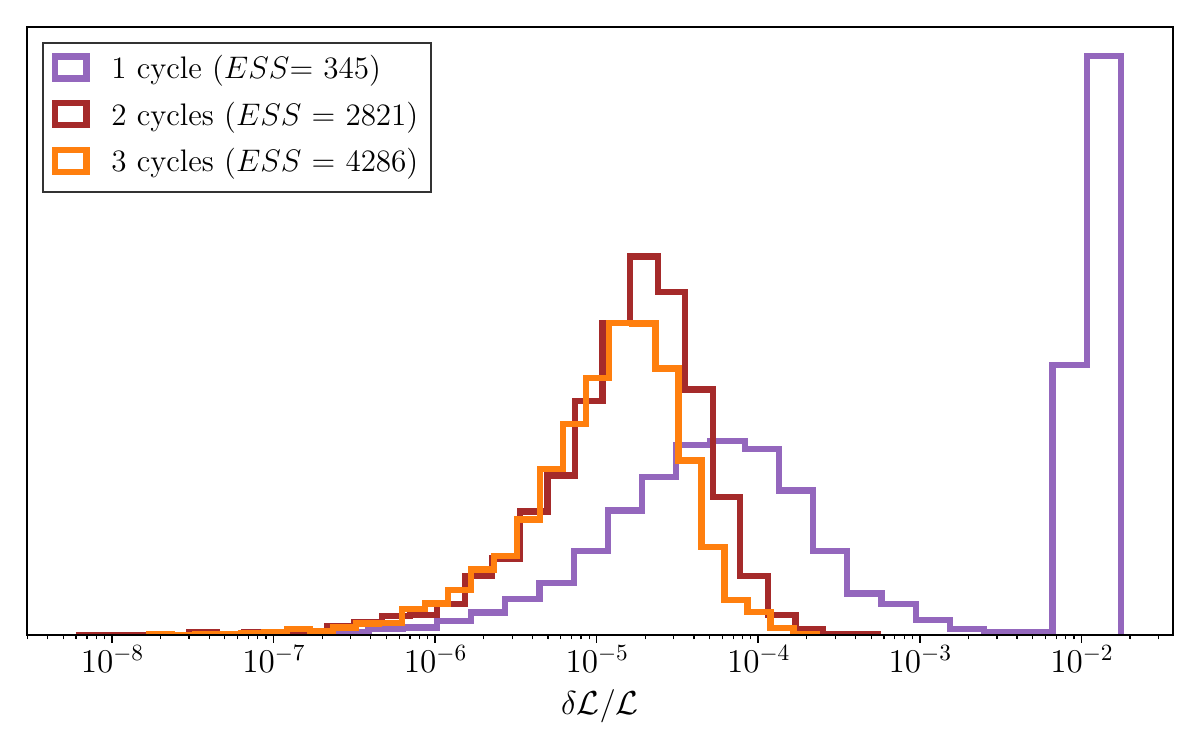}
        }
        \subfigure[]{
            \includegraphics[width=0.47\textwidth]{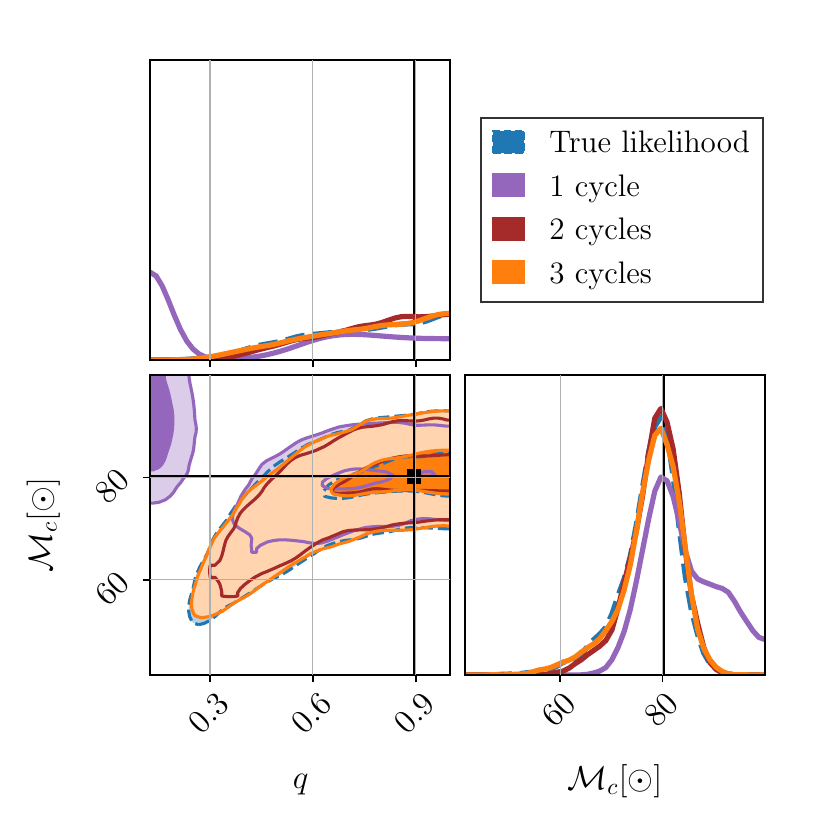}
        }
    
        \caption{Effects of each training cycle on the (partial) posteriors obtained by \texttt{FLEX}. Fig. 4.a shows how the relative error changes between 1, 2, and 3 training cycles, while Fig. 4.b compares the marginal posterior for chirp mass $\mathcal{M}_c$ and mass ratio $q$ between the three cycles of the algorithm, as well as the posterior obtained by sampling the true likelihood. During the first cycle, \texttt{FLEX} incorrectly identifies a spurious peak around a mass ratio of 0.1, alongside the correct one around 0.9. This peak shows up in the error plot as the peak around $10^{-2}$.     }
        \label{fig:retrainingeffects}
        
    \end{figure}

        To further investigate the effect of the retraining cycles, we plot in Fig. \ref{fig:retrainingeffects} an example of the partial posterior PDFs obtained by \texttt{FLEX} at the end of each cycle for an analysis run on a simulated signal with a larger threshold of $ESS/N_{tot} =75\%$ and compare them with the one obtained with the true likelihood, which we consider the ground truth. The histograms of the relative errors for each posterior are shown as well. To quantify the improvement in accuracy between one cycle and the next, we introduce the $ESS$ gain value. If we let $ESS_{i}$ be the Effective Sample Size of the PDF obtained after the $i^{\rm th}$ cycle, we can define the gain of the $i^{\rm th}$ cycle as $G_{i} = ESS_{i}/ESS_{i-1}$. The largest gains are obtained between the first and second cycle, and the median gain of the 97 now active runs is $\overline{G_2} = 6.9^{154.7}_{1.9}$. Fig. \ref{fig:retrainingeffects} offers an example of what is happening: if the network suffers from overfitting, the \texttt{FLEX} NLE will often find a spurious peak in the posterior PDF, which is not present in the original analysis, alongside the real one. Once new training samples are obtained from the spurious peak, the NLE swiftly corrects itself,  resulting in large gains in accuracy. Subsequent cycles will tune the NLE with smaller corrections. Analyzing the following cycles, the median gain has values of $\overline{G_3} = 2.0^{7.9}_{1.4}$ and $\overline{G_4} = 2.1^{4.0}_{1.5}$. We can then expect that every "tuning" cycle improves $ESS$ by a factor $\sim2$.

        \subsubsection{Breakdown of computational costs}
            \label{subsec:compcosts}
            
            To assess the computing and time resources taken by a sampling algorithm, two main metrics can be: 
            
            \begin{itemize}
                \item Number of true likelihood evaluations. It can be used to compare the efficiency of different samplers. 
                \item Total CPU time taken, which indicates the real costs of the algorithm but can vary depending on hardware and the costs of the true likelihood function. 
            \end{itemize}
            
            The previous section gave an overview of the cost in terms of likelihood evaluations, and will be followed up again in section \ref{sec:PE_real_signal} to compare with standard samplers. In this section we instead focus on the second metric and break down its contribution to the total time of the algorithm. and Sec. \ref{subsec:waveform_comparison} will show how the total costs scales with the cost of the true likelihood. To compare numbers across the paper, each \texttt{FLEX} run has been conducted on the same hardware, using only CPUs AMD Rome 7H12.

            \begin{figure}
                \centering
                \includegraphics[width=0.4\textwidth]{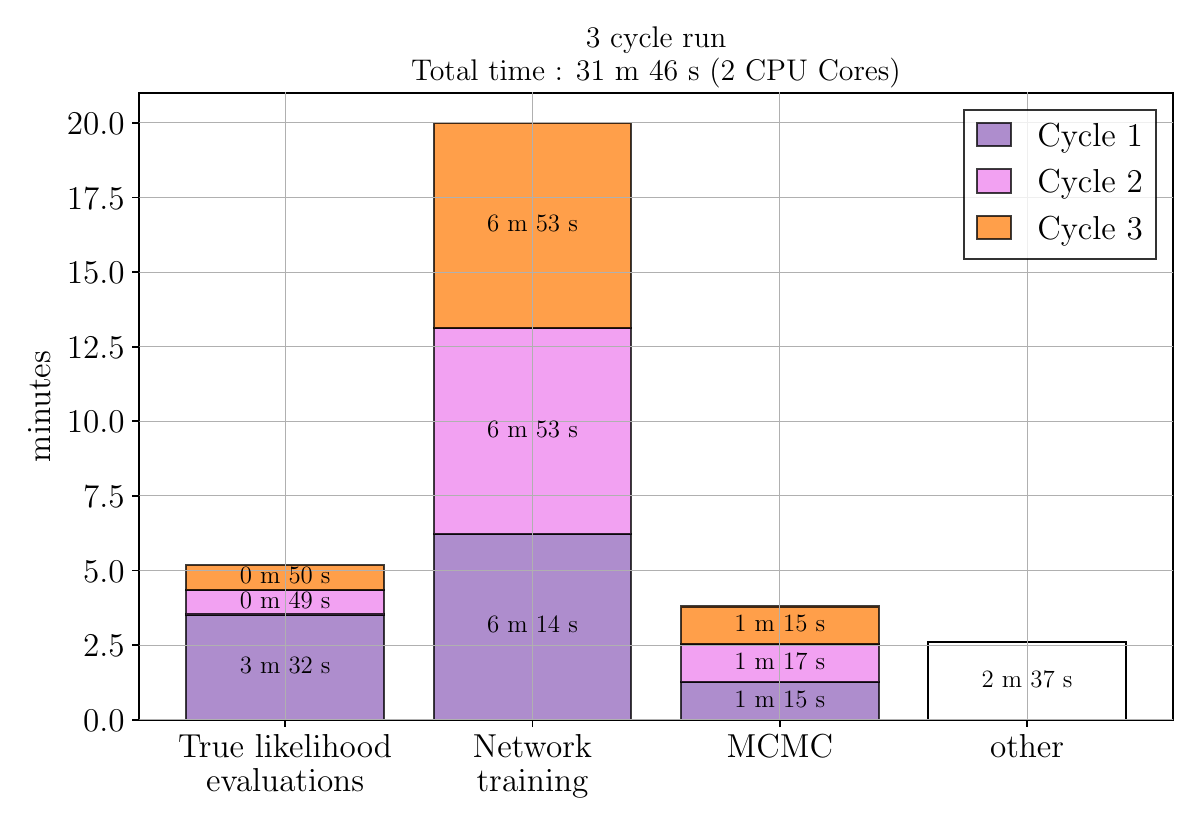}
                \caption{A bar chart showing the distribution of times spent on each phase of the full algorithm for an injected signal with an SNR of 20 and chirp mass of 80 $M_\odot$. The category GW likelihood evaluations represents the time spent by the algorithm in phase 1, category network training represents phase 2, and MCMC represents phase 3. The category "other" encompasses all of the time spent by the plotting scripts and the posterior post-processing. Only the time spent in the 1st phase will scale with the cost of the likelihood, while the time spent by other categories will scale mostly with the quality of hardware. If the posterior is quite hard to sample and presents many non-Gaussian features, the algorithm will take more cycles to converge. The algorithm took 3 cycles to reach convergence for this signal. The architecture at the moment is optimised for reducing the amount of likelihood evaluations, not total time, so the longest portion of time taken by the algorithm is the network training itself.}
                \label{fig:bars}
            \end{figure}

            Fig. \ref{fig:bars} shows a breakdown of the origins of the total computational costs of the algorithm, divided per cycle, from an example run from the simulation set in Sec.~\ref{subsubsec:pp}. Four sources contribute to the total computational costs: true likelihood evaluations, the training of the network, MCMC sampling, and the "other" category, which includes time spent making diagnostic plots and post-processing of results. The run took 3 cycles to converge, and the time spent by each cycle remains fairly constant, except for the computation of true likelihood evaluations for the first cycle, which includes the time needed to run the annealed-KDE algorithm and find the first approximation of the posterior. This is quite an expensive task and requires $10^5$ likelihood evaluations, versus the $2\times 10^4$ required by the subsequent retraining cycles.\\
            
            Since the waveform model used in this study is quite fast, the time spent in the true likelihood phase of each cycle is much less than the time spent on the NLE-related part (Network training and MCMC). For each cycle, the network training takes around 6-7 minutes, while running the full PE algorithm with MCMC on the pre-trained likelihood adds only $\sim$ 1 minute per cycle. The final PE stage requires $\sim 3.5 \times 10^7$ calls to the NLE and shows the real advantage of this approach; excluding the training costs, the single-likelihood evaluation time ($2\mu s$ for the NLE) has been reduced by a factor $\sim 10^3$ with respect to the true approximant ($2ms$ per likelihood). Since the cost of the algorithm related to the NLE will remain unchanged, the factor of speedup is even more pronounced when using more accurate waveform models, and an example of this can be found in Sec. \ref{subsec:waveform_comparison}. \\
            
            The hyperparameters of the algorithm can be modified to further reduce the computational burden. The set used for this injection study is aimed at reducing the number of total likelihood evaluations without compromising accuracy. To reduce total computational costs, better hyperparameters can be used when, like in this scenario, the cost of the single likelihood evaluation is quite low. \\

      \subsection{Parameter estimation on real signals}
        \label{sec:PE_real_signal}

        To further assess the robustness of the algorithm, we tested the \texttt{FLEX} framework on real signals. We have focused on \texttt{GW150914}, detected by an HL-network with a matched-filter SNR of 24 (\cite{LIGOScientific:2016aoc}). This signal allows us to compare results from \texttt{FLEX} to those already publicly released by the LVK collaboration (\cite{LIGOScientific:2019lzm,KAGRA:2023pio,LIGOScientific:2025snk}). The nearly-equal mass and aligned spin nature of the signal additionally means that we can use analytical phase marginalisation.
        As mentioned in Sec. \ref{subsec:followup}, we decided to use the \texttt{pocomc} sampler on the pre-trained NLE to present results of posterior PDFs for this analysis. To ensure convergence across samplers, we raise the threshold to $ESS/N_{post} = 75\%$. We first compare the posterior recovered by \texttt{FLEX} and one recovered by a standard sampler on the true likelihood. To further confirm the robustness of the results, we conclude by analyzing the signal with 4 different waveform models and checking for consistency.

        \subsubsection{Comparison with standard samplers}
        
\begin{figure}
            \centering
            \includegraphics[width=0.47\textwidth]{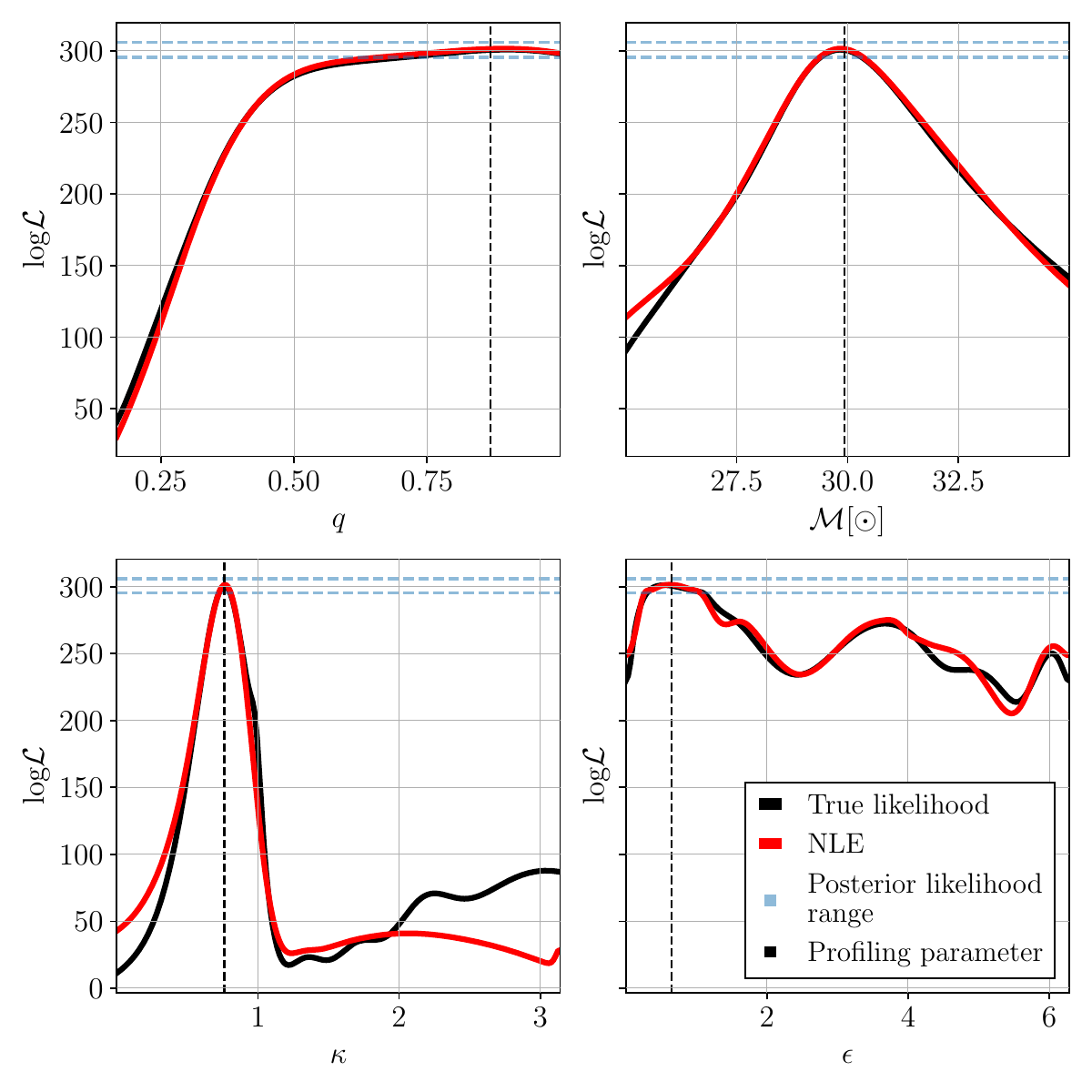}
            \caption{Profiling the true and the \texttt{FLEX} likelihood for \texttt{GW150914} around one point of the posterior (here referenced as the profiling parameter) for mass ratio, chirp mass, azimuth, and zenith angles. The dashed blue lines mark the range of the likelihood values obtained from the posterior samples. The NLE tracks the true likelihood across the entire parameter space, with the largest errors occurring far away from the peak, while in the posterior range, the region where the sampler will spend the most time probing the NLE, the accuracy is high.}
            \label{fig:profilinglikelihood}
        \end{figure}

        Fig.~\ref{fig:gw150914} shows the posterior PDFs obtained by running our algorithm on \texttt{GW150914} and comparing them with the results obtained from \texttt{pocomc}. As before, the analysis was performed with the assumption of aligned spins and the IMRPhenomD waveform model. Following the procedure in \cite{veitch:2015} and \cite{thrane:2019}, the distance parameter, which was initially marginalised can be reconstructed, and the sky-position parameters can be projected from the detector to the geocentric frame of reference as presented in \cite{Romero-Shaw:2020owr}.
        We compute the Jensen-Shannon Divergence (JSD) (\cite{MENENDEZ1997307}) between the two distributions to quantify the difference between the 1D marginalised posteriors and report them in Fig.~\ref{fig:gw150914}. Following the procedure in \cite{ashton:2021}, for our sample size of a few thousand, the two distributions can be considered statistically equivalent if their JSD lies below 1.5 millinats. This threshold is met for every parameter. While the bulk of the posterior is well represented by \texttt{FLEX}, there are some small visual differences regarding the secondary sky-position mode. Addressing the capabilities of \texttt{FLEX} for characterising multimodal posteriors is one of the main challenges for future development. To further examine the behaviour of the NLE, we profiled it against the true likelihood in Fig. \ref{fig:profilinglikelihood} around a random posterior sample. The blue lines indicate the range in which the posterior likelihood values lie. As we can see, the NLE approximates very closely the neural likelihood in the region around the peak, while allowing for larger errors in low-likelihood regions. This behaviour is there by design: by sampling most of the training set around the region of parameter space where the likelihood is highest, the network will also be more accurate, and in low-likelihood regions, the NLE does not need to reach a high level of accuracy to still give acceptable results.\\

        The computational costs for both algorithms are reported in Tab \ref{tab:compcosts}.  The total number of true likelihood evaluations for the standard Bayesian inference is $1.3\times 10^7$, while the \texttt{FLEX} implementation only evaluates the gravitational wave likelihood during the generation of the training set, $1.8\times10^5$ times, and the sampler only calls the very cheap to evaluate NLE: \texttt{FLEX} is able to reduce the number of true likelihood calls by 98.6\% without compromising the accuracy. Considering the extra time used by the other parts of the algorithm and computing the CPU wall time, \texttt{FLEX} required 20 times less computational resources with respect to the standard algorithm. The log-evidence value obtained by the two methods differs by 1 unit. This error is larger than the one introduced by changing samplers between \texttt{dynesty} and \texttt{pocomc}. Since the posteriors visually agree for all parameters, we suspect that this difference is driven by errors in modeling the tails of the distribution. In future work, we plan to explore and better characterise this issue.  
            \begin{figure}
                \centering
                \includegraphics[width=0.47\textwidth]{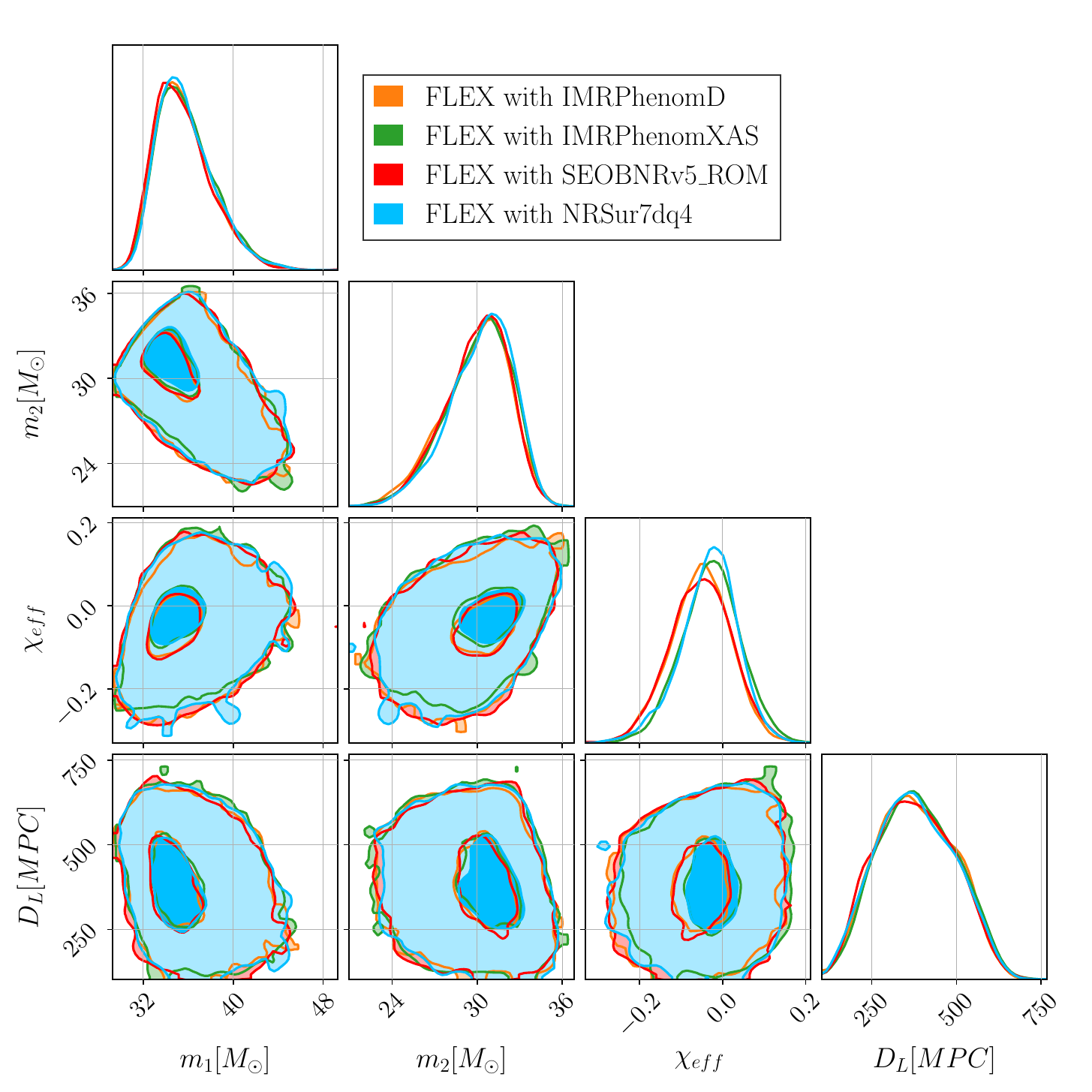}
                \caption{ \texttt{GW150914} posteriors recovered with different waveforms with the \texttt{FLEX} algorithm. Only luminosity distance and intrinsic parameters projected to the source frame are reported. The results are consistent between the different analyses, as expected from previous literature on the subject. The small difference encountered in $\chi_{eff}$ between SEOB and PhenomD with respect to PhenomXAS and NRSur is also reported in previous works and explained in Sec.~\ref{subsec:waveform_comparison}. 
                }
                \label{fig:gw150914waveforms}
            \end{figure}

        \subsubsection{Waveform comparison}
            \label{subsec:waveform_comparison}
            When comparing different waveform models, \texttt{GW150914} lies in a well-studied region of parameter space for most of the waveform models, and posterior PDFs recovered by different models are expected to agree. To further validate \texttt{FLEX} and predict the costs of the algorithm when analysing with more expensive likelihood functions, we performed the analyses with 4 waveforms. We choose two phenomenological models, IMRPhenomD, IMRPhenomXAS (\cite{pratten:2020}), which are fast aligned-spin approximants, the effective one body SEOBNRv5\_ROM (\cite{pompili:2023}) approximant, sped up with the use of Reduced Order Modes (ROMs) \cite{cotesta:2020}, and the numerical relativity surrogate model NRSur7dq4 \cite{varma:2019}, limited to the leading-order (2,2) mode to allow for analytical phase marginalisation. Each analysis was run independently, with the neural network and training samples being randomly initialised each time. The recovery of intrinsic parameters for the different \texttt{FLEX} analysis is reported in Fig. \ref{fig:gw150914waveforms} and detailed timing, likelihood, and log evidence values are reported in Tab. \ref{tab:compcosts}. All of the analyses return statistically consistent results, as well as comparable values of log evidence. The small bias between the values of $\chi_{eff}$ recovered by different models is a known phenomenon, and it is reported in Fig. 19 of Ref.~\cite{pratten:2020} as well as Fig. 23 of Ref.~\cite{pompili:2023}. This indicates that the relative errors introduced by approximating the likelihood function with \texttt{FLEX} are smaller than the waveform systematic errors. This ensures that \texttt{FLEX} is accurate enough to perform model-comparison analysis.

            \begin{table*}
                \centering
                \begin{tabular}{|l|l|l|m{5.8em}|m{3em}|l|l|m{5.5em}|}
                    \hline
                    \textbf{Waveform} & \textbf{Sampler}& \textbf{$\log{\mathcal{Z}}$} & \textbf{No. waveform evaluations}& \textbf{No. of Cycles}&\textbf{Total CPU time} & \textbf{Time per cycle}& \textbf{\%Waveform evaluation time}  \\
                    \hline
                    IMRPhenomD &\texttt{pocomc}& 286.1 & $1.3\times10^7$& - & 656m & - & - \\
                    \hline
                    IMRPhenomD & \texttt{Dynesty}  & 286.0 & $2.3\times10^7$& - & 1008m & - & - \\
                    \hline
                    IMRPhenomD & \texttt{FLEX}+\texttt{pocomc} &285.2 & $1.8\times10^5$ & 5 & 40m & 7m29s &16\% \\
                    \hline
                    IMRPhenomXAS & \texttt{FLEX}+\texttt{pocomc}& 284.7 & $1.6\times10^5$& 4 &31m  & 7m9s &20\% \\
                    \hline
                    SEOBNRv5\_ROM & \texttt{FLEX}+\texttt{pocomc}& 284.5 & $1.4\times10^5$& 3&38m & 11m31s &42\% \\
                    \hline
                    NRSur7dq4 & \texttt{FLEX}+\texttt{pocomc}& 284.3 & $1.8\times10^5$& 4&109m &19m52s &70\% \\
                    \hline
                \end{tabular}
                \caption{Bayesian evidence and computational costs for different analyses on \texttt{GW150914}. All have been performed with aligned spins. Two standard analyses run with \texttt{pocomc} and \texttt{dynesty} samplers on the IMRPhenomD waveform model are compared with 4 different analyses with the \texttt{FLEX} framework using different waveform models with varying computational costs. Results in the first two rows have been obtained using the true likelihood in a standard analysis.}
                \label{tab:compcosts}
            \end{table*}
            
        \begin{figure}
            \centering
            \includegraphics[width=0.47\textwidth]{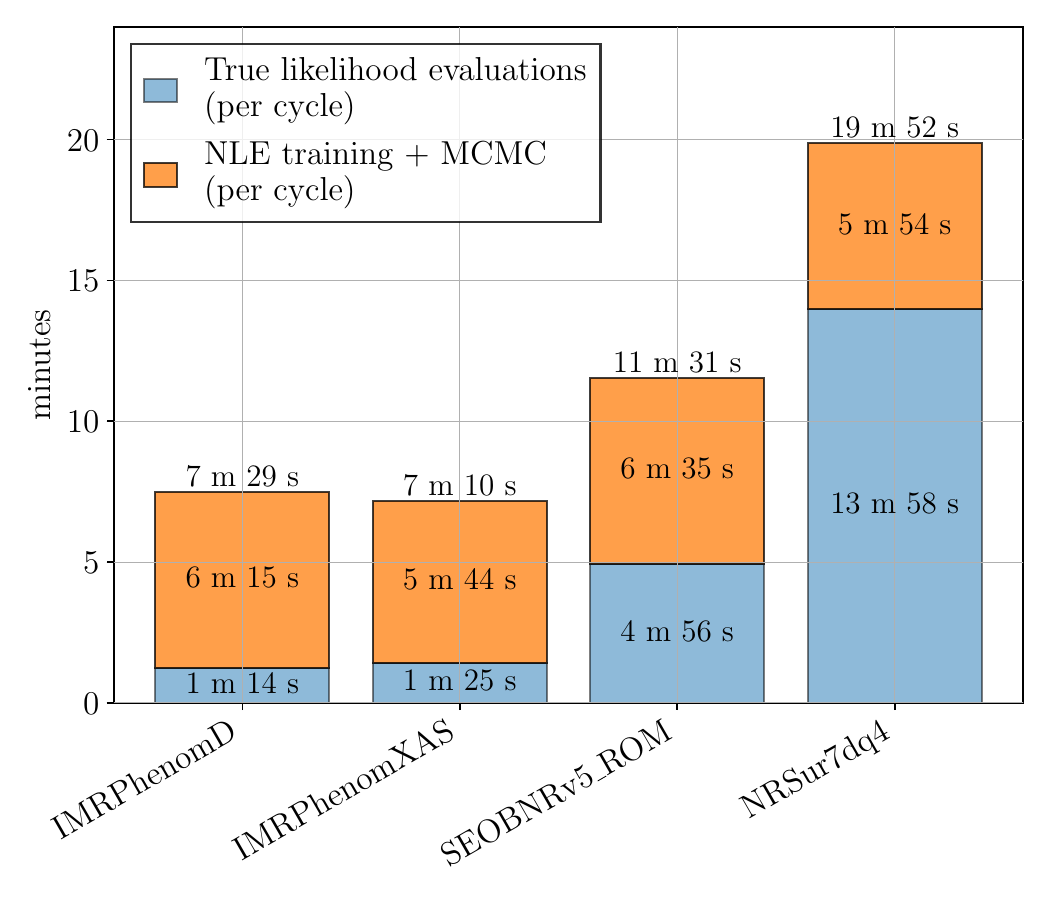}
                \caption{Mean time per cycle taken by the different waveform analysis runs on \texttt{GW150914}. For non-expensive waveforms like the ones from the Pheonom family, the NLE training and the MCMC take the longest time, while for more expensive waveforms like NRSur, this time is a smaller percentage. 
                \label{fig:gw150914_time_per_cycle}}
        \end{figure}
        
        \begin{figure*}
            \centering
            \includegraphics[width=0.95\textwidth]{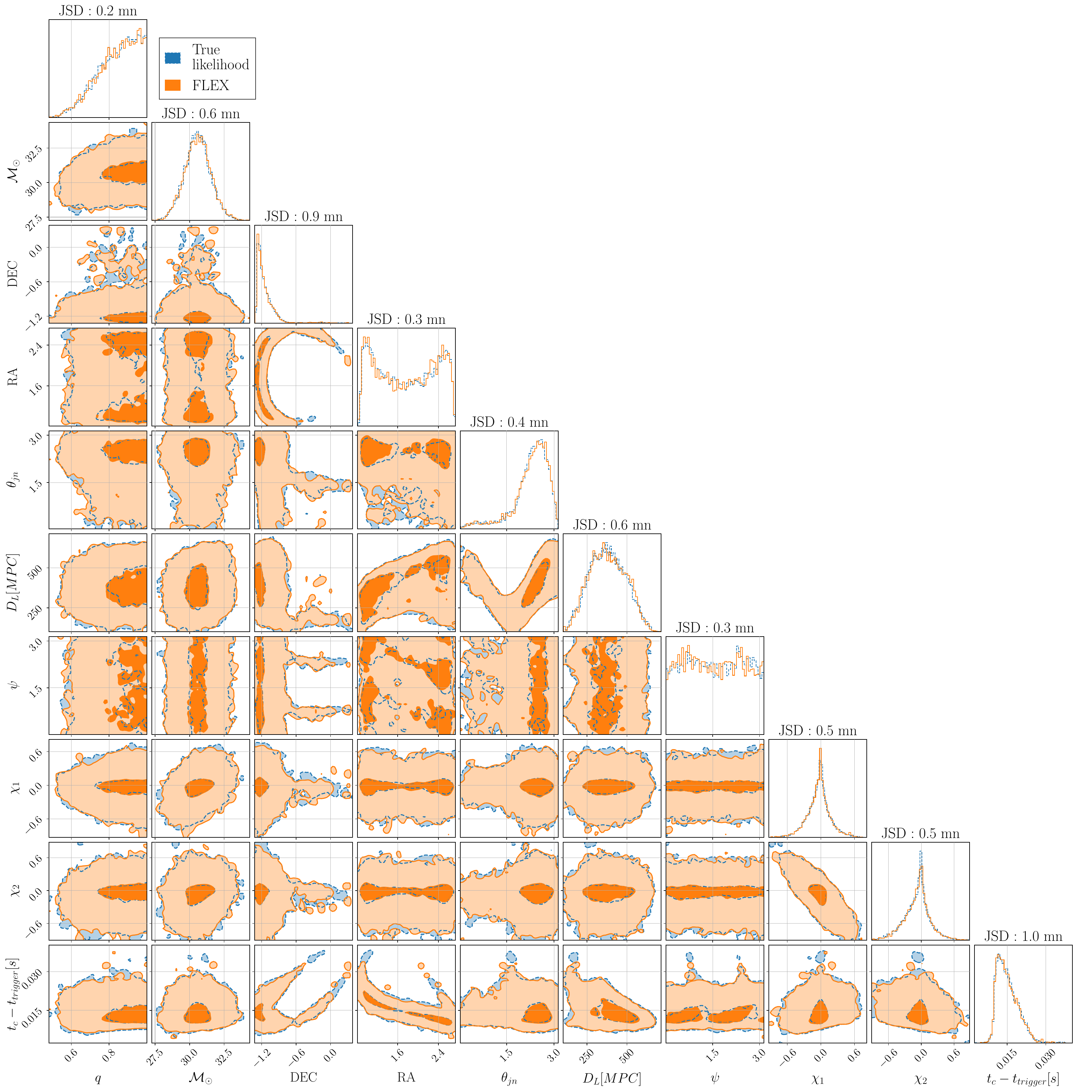}
                \caption{Comparison between posteriors for \texttt{GW150914} obtained by sampling the true and the \texttt{FLEX} neural likelihood. The PDFs visually agree for all parameters. The 1-dimensional Jensen-Shannon Divergence (JSD) between the marginalised posteriors for each parameter is below 1.5 millinats, meaning that the two posterior draws are statistically consistent with each other. The two posteriors have been obtained by running on the same hardware. \texttt{pocomc} on the vanilla likelihood required 1.3$\times10^7$ likelihood evaluations, while building the \texttt{FLEX} NLE only required 1.8$\times 10^5$.}
            \label{fig:gw150914}
        \end{figure*}

\section{Discussion }
\label{sec:discussion}

In the past sections, we have shown that our developed method, \texttt{FLEX}, shows considerable promise, and at the same time comes with some caveats. Below we outline some possible improvements to our algorithm as well as long-term aims to make our method more streamlined.

\begin{itemize}

    \item \textbf{Obtaining training samples}: All results presented in this study used a weighted KDE approach to obtain training samples; however, we have tried optimisers such as \emph{Differential Evolution}, which also yield comparable results. The requirement of the training samples approximating the posterior distribution closely leaves room for further development in this direction. 
    \item \textbf{Multimodal distributions}: The quality of the approximate likelihood is limited by how well the initial training samples resemble the final posterior. This affects parameters with known multimodal posteriors—particularly extrinsic ones such as sky location and the polarization angle $\psi$. While working in the alternate parameterization (\emph{c.f.} Sec.~\ref{subsec:params}) helps in sampling, these parameters could be poorly constrained.

    \item \textbf{Dimensionality constraints}: To reduce computational cost, we marginalise over the luminosity distance and phase and restrict to aligned-spin systems. For precessing sources and higher modes waveform models, where phase marginalization is not possible, the addition of extra parameters exacerbates multimodality and slows down convergence. Also, full parameter estimation considers detector calibration uncertainties. The likelihood now depends on an extra 20 parameters per detector, and such a high-dimensional space can be hard to tackle for machine learning algorithms. 

    \item \textbf{High-SNR regime}: As with many samplers, performance degrades for high SNR signals. We have validated our algorithm up to SNRs $\sim 40$, but going forward, and particularly with improving detector sensitivities, we expect louder signals frequently. We will address this challenge in future work.
\end{itemize}

\section{Conclusions}
\label{sec:conlcusions}

In this work, we have applied a neural likelihood estimator to real GW data for the first time. Our algorithm combines the benefits of flexibility, ease of implementation, and minimal hardware requirements. Because the neural network is trained on-the-fly, the method is highly adaptable to different waveform models, priors, and samplers, with no need for expensive pre-training.

Compared to standard analysis, our approach achieves significant speed-ups, reducing both the number of true likelihood evaluations and the wall-clock time required for parameter estimation. On average, our method requires 100 times fewer true likelihood evaluations and at least 10 times lower computational costs.
The method remains robust up to moderate signal-to-noise ratios and high mass BBH events. Additionally, our algorithm has the unique feature of providing us with a full approximation of the true likelihood, which can be used as a drop-in replacement for subsequent analysis. We test this proof-of-concept by re-using the NLE in another sampler that allows us to calculate the Bayesian evidence.\\

Looking forward, we see several promising directions for extending this work. Our preliminary studies suggest that the method is applicable when adding further parameters in the waveform model; a broader investigation into analysing signals with more parameters, as well as improvements to the network, is left for a future publication. Currently, each new signal requires retraining the neural likelihood from scratch. For a fixed waveform parameterisation, transfer learning may substantially accelerate inference on different signals and is also left for a future study. The NLE shows promise to reduce the cost of follow-up analysis even further by fine-tuning the neural network to approximate likelihood functions closely resembling the original one. Future improvements will target better handling of high-dimensional and multimodal posteriors, as well as integrating different detector networks. Taken together, these features make neural likelihood estimation a compelling, scalable tool for fast and flexible GW parameter inference.

\section{Acknowledgments}
The authors would like to thank Michael Williams and Thibeau Wouters for the insightful comments. We also thank Michael Williams for help with setting up the pocomc sampler and Anna Puecher for her help with the \texttt{Eryn} sampler. The authors thank the SURF and the National Supercomputer Snellius in the Netherlands, through the computing grant EINF-12633, and the LIGO laboratory data grid clusters at CIT supported by the NSF grant PHY-0757058 and PHY-0823459. This material is based upon work supported by NSF's LIGO Laboratory, which is a major facility fully funded by the National Science Foundation. This research has made use of data or software obtained from the Gravitational Wave Open Science Center (gwosc.org), a service of the LIGO Scientific Collaboration, the Virgo Collaboration, and KAGRA. This material is based upon work supported by NSF's LIGO Laboratory which is a major facility fully funded by the National Science Foundation, as well as the Science and Technology Facilities Council (STFC) of the United Kingdom, the Max-Planck-Society (MPS), and the State of Niedersachsen/Germany for support of the construction of Advanced LIGO and construction and operation of the GEO600 detector. Additional support for Advanced LIGO was provided by the Australian Research Council. Virgo is funded, through the European Gravitational Observatory (EGO), by the French Centre National de Recherche Scientifique (CNRS), the Italian Istituto Nazionale di Fisica Nucleare (INFN) and the Dutch Nikhef, with contributions by institutions from Belgium, Germany, Greece, Hungary, Ireland, Japan, Monaco, Poland, Portugal, Spain. KAGRA is supported by Ministry of Education, Culture, Sports, Science and Technology (MEXT), Japan Society for the Promotion of Science (JSPS) in Japan; National Research Foundation (NRF) and Ministry of Science and ICT (MSIT) in Korea; Academia Sinica (AS) and National Science and Technology Council (NSTC) in Taiwan. The authors would like to thank the LIGO, Virgo and KAGRA (LVK) collaborations for providing software that was intrumental for this work like the gravitational wave open science center (GWOSC) and the \texttt{bilby} package.

\section{Data Availability}
Real GW data used in this paper is publicly available in \texttt{GWOSC}. The rest of the analysis developed in this paper will be provided by the authors upon reasonable request. 

\bibliographystyle{mnras}
\bibliography{references}

\end{document}